\documentclass[twocolumn,floats,floatfix,superscriptaddress,aps,pra]{revtex4}
\usepackage{amsfonts}
\usepackage{amssymb}
\usepackage{amsmath}
\usepackage{calc}
\usepackage{graphicx}
\usepackage{bm}

\def\be{ \begin{equation} }
\def\ee{ \end{equation} }
\def\bea{ \begin{eqnarray} }
\def\eea{ \end{eqnarray} }
\def\bse{ \begin{subequations} }
\def\ese{ \end{subequations} }
\def\i{i}
\def\e{\,\text{e}}

\def\U{\mathbf{U}}
\def\A{\mathcal{A}}

\def\d{\text{d}}
\def\T{\mathcal{T}}
\def\T{T}

\def\half{\tfrac12}

\def\quarter{\tfrac14}

\def\sec{\section}

\begin{document}

\author{Boyan T. Torosov}
\affiliation{Institute of Solid State Physics, Bulgarian Academy of Sciences, 72 Tsarigradsko chauss\'{e}e, 1784 Sofia, Bulgaria}
\author{Nikolay V. Vitanov}
\affiliation{Department of Physics, St Kliment Ohridski University of Sofia, 5 James Bourchier blvd, 1164 Sofia, Bulgaria}


\title{Arbitrarily accurate twin composite $\pi$ pulse sequences}

\date{\today}

\begin{abstract}
We present three classes of symmetric broadband composite pulse sequences. 
The composite phases are given by analytic formulas (rational fractions of $\pi$) valid for any number of constituent pulses.
The transition probability is expressed by simple analytic formulas and the order of pulse area error compensation grows linearly with the number of pulses.
Therefore, any desired compensation order can be produced by an appropriate composite sequence; in this sense, they are arbitrarily accurate.
These composite pulses perform equally well or better than previously published ones.
Moreover, the current sequences are more flexible as they allow total pulse areas of arbitrary integer multiples of $\pi$.
\end{abstract}

\maketitle


\sec{Introduction}

Among the coherent control techniques in quantum physics, composite pulses (CPs) have the unique advantage of combining
  ultrahigh accuracy similar to resonant techniques with robustness to parameter imperfections similar to adiabatic passage techniques.
Although the required total pulse area of a CP is typically a few times larger than in the resonance techniques, it is still significantly less than the typical pulse areas in the adiabatic techniques.

CPs have been introduced and used extensively in nuclear magnetic resonance (NMR) \cite{Levitt1979,Freeman1980,Levitt1986,Freeman1997}.
In fact, composite polarisation plates, which are based on the same mathematical principles, have been known in polarisation optics much earlier \cite{West49,Destriau49,Pancharatnam55,Harris64,McIntyre68}.
Recently, CPs have found numerous applications in quantum information \cite{Gulde2003,Schmidt-Kaler2003,Timoney2008,Monz2009,Ivanov2011,Ivanov2015},
 quantum optics \cite{Torosov2011PRA,Torosov2011PRL,Schraft2013,Genov2014PRL}, and even frequency conversion in classical optics \cite{Genov2014JOpt,Rangelov2014}.

The composite pulse sequence is a finite train of pulses with well defined relative phases between them.
These phases are control parameters: they are determined by the desired excitation profile.
In general, CPs allow for great flexibility to shape the excitation profile in essentially any desired manner, an objective which is impossible with a single resonant pulse or adiabatic techniques.
In particular, broadband (BB) CPs can cancel imperfections of a single pulse due to deviations in the pulse area, frequency offset, chirp, etc.
Alternatively, one can create narrowband CPs \cite{Tycko1984,Tycko1985,Wimperis1994,Shaka1984,Ivanov2011OL,Vitanov2011}, which enhance the single pulse sensitivity to variations in a certain parameter, which can be used, for instance, for localisation or sensing.
Passband CPs \cite{Wimperis1994,Kyoseva2013} combine the features of broadband and narrowband CPs: highly efficient excitation within a certain range of an interaction parameter and negligibly small excitation outside.

In this paper, we present three sets of BB composite $\pi$ pulses, which compensate imperfections in the peak Rabi frequencies and the pulse durations.
The composite phases are given by simple analytic formulas --- rational multiples of $\pi$.
The transition probability for each set is given by a simple analytic formula too, which makes it possible to explicitly assess the high-efficiency range.
We prove that these CPs perform equally well or better than previously known CPs.
Moreover, they are more flexible because the total nominal pulse area can be any integer multiple of $\pi$, whereas previously known BB CPs used pulse areas given by $(2n+1)\pi$ \cite{Torosov2011PRA} or $2^n \pi$ \cite{Levitt82,Levitt83}, where $n$ is an integer.
Our composite sequences can contain arbitrarily many pulses (with a total nominal pulse area $n\pi$) and are accurate up to order $O(\epsilon^{2n})$, where $\epsilon$ is the pulse area error; hence they can be made accurate to any order in $\epsilon$.


\sec{Twin composite pulse sequences}
We begin by a brief introduction to the theory of composite pulses. A general SU(2) propagator is parameterized using the complex Cayley-Klein parameters $a$ and $b$ ($|a|^2+|b|^2=1$) as
\be
\U = \left[\begin{array}{cc}
a & b \\
 - b^\ast & a^\ast
\end{array}\right] .
\ee
A constant phase shift in the Rabi frequency $\Omega\to\Omega\e^{\i\phi}$ is imprinted into the propagator as
\be
\U_\phi = \left[\begin{array}{cc}
a & b \e^{\i\phi}\\
 - b^\ast \e^{-\i\phi} & a^\ast
\end{array}\right] .
\ee
For resonant excitation ($\Delta=0$), which we assume, $a=\cos(\A/2)$ and $b=-\i\sin(\A/2)$, where $\A=\int_{t_i}^{t_f}\Omega(t)\d t$ is the pulse area.
The transition probability in this case is $P_{1\to 2} = \sin^2(\A/2)$.
It is equal to 1 for a resonant $\pi$ pulse, i.e. for $\A=\pi$.
It is accurate up to order $O(\epsilon^2)$ to deviations of the pulse area, $\A=\pi(1+\epsilon)$, from this perfect value: $P_{1\to 2} = 1-\pi^2\epsilon^2/4 +\ldots$.
By replacing the single $\pi$ pulse by a CP one can enhance the robustness to pulse area errors up to order $O(\epsilon^{2n})$, with $n>2$.

The propagator of the composite pulse sequence,
\be\label{SN}
S_N = (\A_1)_{\phi_1} (\A_2)_{\phi_{2}} \cdots (\A_{N-1})_{\phi_{N-1}} (\A_{N})_{\phi_{N}} ,
\ee
is the product of phase shifted propagators,
\be\label{Utot}
\U^{(N)} = \U_{\phi_N}(\A_N)\U_{\phi_{N-1}}(\A_{N-1})\cdots \U_{\phi_{2}}(\A_{2}) \U_{\phi_{1}}(\A_{1}) ,
\ee
where $\A_j$ and $\phi_j$ are the pulse area and the phase of pulse $j$ $(j=1,2,\ldots, N)$.
The objective is to choose $\A_j$ and $\phi_j$ in such way that to produce a BB excitation profile to the highest possible order in the deviation $\epsilon$.

We build our CPs as twin sequences,
\be\label{UCP}
\T_N = S_N \widetilde{S}_N ,
\ee
where $\widetilde{S}_N$ is the inverted sequence of $S_N$ of Eq.~\eqref{SN},
\be
\widetilde{S}_N = (\A_N)_{\phi_N} (\A_{N-1})_{\phi_{N-1}} \cdots (\A_2)_{\phi_{2}} (\A_1)_{\phi_{1}} .
\ee
Hence, if $S_N$ has $N$ pulses, the total BB sequence $\T_N$ contains $2N-1$ pulses because the two adjacent pulses $(\A_{N})_{\phi_{N}} (\A_{N})_{\phi_{N}}$ are merged into one pulse $(2\A_{N})_{\phi_{N}}$.

We consider three sets of composite sequences, which we label as $T_N^{(1)}$, $T_N^{(2)}$, and $T_N^{(3)}$.
They are composed of pulses of areas
\be
A = \frac{\pi}{2} (1+\epsilon),\quad
B = {\pi} (1+\epsilon),\quad
C = 2{\pi} (1+\epsilon).
\ee
These are nominal (i.e., for zero error, $\epsilon=0$) $\pi/2$, $\pi$ and $2\pi$ pulses.

\subsection{Type-1 twin pulses}

The first type of twin $\pi$ pulse sequences have the form  $\T_N^{(1)} = S_N^{(1)} \widetilde{S}_N^{(1)}$, with
\be\label{T1}
S_N^{(1)} = A_{\phi_1} B_{\phi_2} B_{\phi_3} \cdots  B_{\phi_{N-1}} A_{\phi_N}.
\ee
Hence the first and last pulses are nominal $\pi/2$ pulses and all pulses in between are nominal $\pi$ pulses.
The composite phases are given by the simple analytic formula
\be\label{phasesCP1}
 \phi_k = \frac{(k-1)^2\pi}{2(N-1)} \quad (k=1,2,\ldots , N).
\ee
Explicitly, the first few twin sequences read
\bse
\begin{align}
\T_2^{(1)}&= A_0 B_{\frac12\pi} A_0, \label{T1-2}\\
\T_3^{(1)}&= A_0 B_{\frac14\pi} B_{\pi} B_{\frac14\pi} A_0, \label{T1-3}\\
\T_4^{(1)}&= A_0 B_{\frac16\pi} B_{\frac23\pi} B_{\frac32\pi} B_{\frac23\pi} B_{\frac16\pi} A_0, \label{T1-4}\\
\T_5^{(1)}&= A_0 B_{\frac18\pi} B_{\frac12\pi} B_{\frac98\pi} B_0 B_{\frac98\pi} B_{\frac12\pi} B_{\frac18\pi} A_0, \label{T1-5}
\end{align}
\ese
where we have used that $A_\phi A_\phi = B_\phi$ when merging $S_N^{(1)}$ and $\widetilde{S}_N^{(1)}$.
The total pulse area for the sequence $\T_N$ is $2(N-1)\pi$.

The proof that the phases \eqref{phasesCP1} lead to BB excitation profiles is too cumbersome to be derived in full; a sketch is presented in the Appendix.

The transition probability for an $N$-pulse sequence of this type reads
\be\label{T1-p}
P_N^{(1)} = 1 - \sin^{4(N-1)}\left( \tfrac12{\pi\epsilon} \right) ,
\ee
Obviously, it is accurate up to order $O(\epsilon^{4(N-1)})$.

\begin{figure}[tb]
\includegraphics[width=0.9\columnwidth]{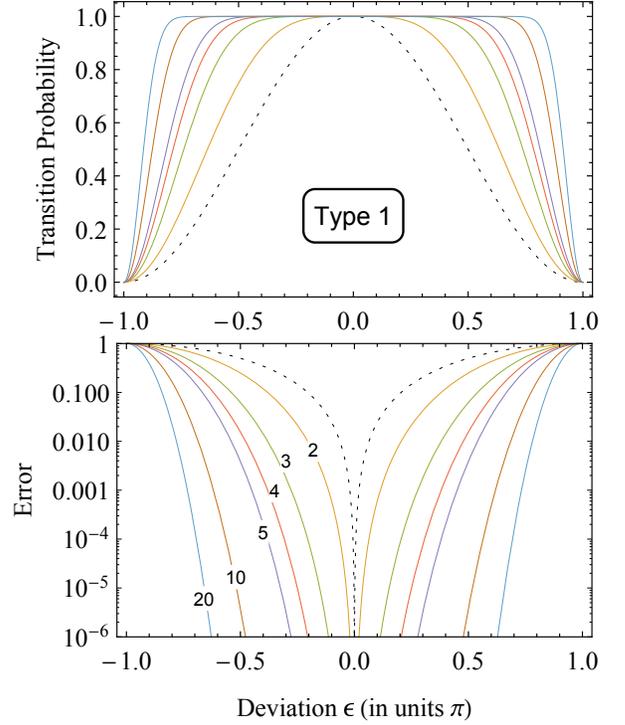}
\caption{(Top) Excitation profiles $P_{N}^{(1)}$ for the BB composite sequences of type-1 for $N=2,3,4,5,10,20$. (Bottom) Transition probability error $1-P_{N}^{(1)}$ in log scale. The dashed curve depicts the single-pulse $\sin^2$ profile for comparison.}
\label{fig1}
\end{figure}

The performance of these sequences is shown in Fig.~\ref{fig1}.
The longer the CPs, the higher the compensation order and the broader the excitation profiles.

\subsection{Type-2 twin pulses}

The second type of twin $\pi$ pulse sequences have the form  $\T_N^{(2)} = S_N^{(2)} \widetilde{S}_N^{(2)}$, with
\be\label{T2}
S_N^{(2)} = A_{\phi_1} B_{\phi_2} B_{\phi_3} \cdots  B_{\phi_{N-1}} B_{\phi_N},
\ee
with the composite phases
\be\label{phasesCP2}
 \phi_k = \frac{2(k-1)^2\pi}{2N-1} \quad (k=1,2,\ldots , N).
\ee
Explicitly, for the first few sequences we have
\bse
\begin{align}
\T_2^{(2)} &= A_0 C_{\frac23\pi} A_0,\label{T2-2} \\
\T_3^{(2)} &= A_0 B_{\frac25\pi} C_{\frac85\pi} B_{\frac25\pi} A_0,\label{T2-3} \\
\T_4^{(2)} &= A_0 B_{\frac27\pi} B_{\frac87\pi} C_{\frac47\pi} B_{\frac87\pi} B_{\frac27\pi} A_0,\label{T2-4} \\
\T_5^{(2)} &= A_0 B_{\frac29\pi} B_{\frac89\pi} B_{0} C_{\frac{14}9\pi} B_{0} B_{\frac89\pi} B_{\frac29\pi} A_0,\label{T2-5}
\end{align}
\ese
where we have used that $B_\phi B_\phi=C_\phi$.
The total pulse area for these sequences is $(2N-1)\pi$ and the transition probability is
\be\label{T2-p}
P_N^{(2)} = 1 - \sin^{4N-2}\left( \tfrac12{\pi\epsilon} \right) .
\ee

In Fig.~\ref{fig2} we illustrate the excitation profiles of these sequences.
Clearly, the larger is $N$, the broader is the excitation profile.

\subsection{Type-3 twin pulses}

\begin{figure}[tb]
\includegraphics[width=0.9\columnwidth]{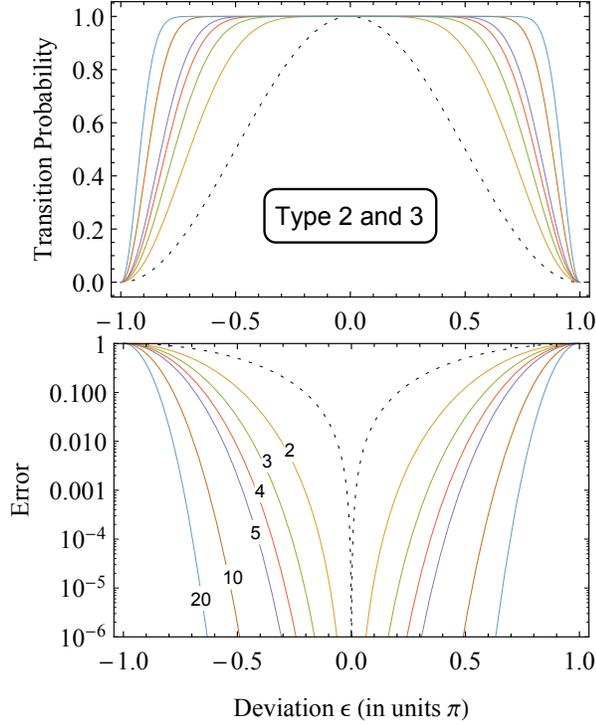}
\caption{(Top) Excitation profiles $P_{N}^{(2,3)}$ for the BB composite sequences of types 2 and 3 for $N=2,3,4,5,10,20$.
(Bottom) Transition error $1-P_{N}^{(2,3)}$ in log scale. The dashed curve depicts the single-pulse $\sin^2$ profile for comparison.}
\label{fig2}
\end{figure}

The third type of twin $\pi$ pulse sequences have the form $\T_N^{(3)} = S_N^{(3)} \widetilde{S}_N^{(3)}$ where
\be\label{T3}
S_N^{(3)} = \widetilde{S}_N^{(2)} = B_{\phi_N} B_{\phi_{N-1}} \cdots  B_{\phi_2} A_{\phi_1},
\ee
with the same composite phases \eqref{phasesCP2} as the $S_N^{(2)}$ sequences but in the reverse order.
Therefore, these sequences can be written also as $\T_N^{(3)} = \widetilde{S}_N^{(2)} S_N^{(2)}$.
The first few sequences are
\bse
\begin{align}
\T_2^{(3)} &= B_0 B_{\frac23\pi} B_0,\label{T3-2} \\
\T_3^{(3)} &= B_0 B_{\frac25\pi} B_{\frac85\pi} B_{\frac25\pi} B_0,\label{T3-3} \\
\T_4^{(3)} &= B_0 B_{\frac27\pi} B_{\frac87\pi} B_{\frac47\pi} B_{\frac87\pi} B_{\frac27\pi} B_0,\label{T3-4} \\
\T_5^{(3)} &= B_0 B_{\frac29\pi} B_{\frac89\pi} B_{0} B_{\frac{14}9\pi} B_{0} B_{\frac89\pi} B_{\frac29\pi} B_0.\label{T3-5}
\end{align}
\ese
The total pulse area and the transition probability $P_N^{(3)}$ are the same as for the $\T_N^{(2)}$ sequence, Eq.~\eqref{T2-p}.
Hence Fig.~\ref{fig2} illustrates the excitation profiles of the sequences of type-3 too.
Like type-2, the type-3 CPs are accurate up to order $O(\epsilon^{4N-2})$.

\sec{Comparison with other composite $\pi$ pulses}

In this section we compare the performance of the present BB sequences with other CPs known in the literature.

The total pulse area of our CPs is $2(N-1)\pi$ for type 1 and $(2N-1)\pi$ for types 2 and 3.
The BB sequences, derived in \cite{Torosov2011PRA}, have a total pulse area of odd multiples of $\pi$.
Hence, we have a greater flexibility in the total pulse area, which can be an odd or an even multiple of $\pi$.

For a given total pulse area $A_{\text{tot}} = (2N-1)\pi$ or $A_{\text{tot}} = 2(N-1)\pi$, our CPs deliver compensation to pulse area errors up to order $O(\epsilon^{2A_{\text{tot}}/\pi})$ --- the same as for the CPs in \cite{Torosov2011PRA}.
In \cite{Torosov2011PRA} it was shown that these profiles outperform the profiles, emerging from some widely used composite sequences \cite{Wimperis1994,Tycko1985,Brown}.
Hence, this is also valid for the current CPs.

\begin{figure}[tb]
\includegraphics[width=8.5cm]{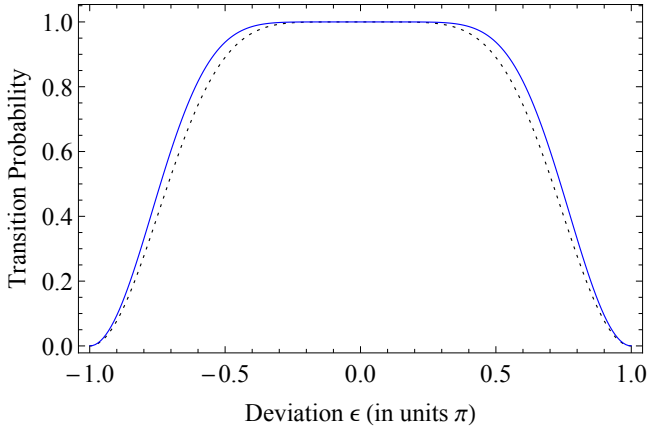}
\caption{Comparison between type-1 twin pulse \eqref{T1-3}, $L_3$ \eqref{L3}, and $L_4$ \eqref{L4}. The solid blue line shows the profile of the type-1 sequence for $N=3$. The dotted black line corresponds to the nine-pulse sequence $L_3$. The profile of $L_4$ coincides with the type-1 profile.}
\label{fig3}
\end{figure}

We now focus on the comparison of our pulses with those of Refs.~\cite{Levitt82,Levitt83}.
To be specific, we will consider four of these sequences, which we call $L_1$, $L_2$, $L_3$ and $L_4$.
Translated from the NMR language, they are written as
\bse
\begin{align}
L_1 &=  A_{0}B_{\frac12\pi}A_{0} ,\\
L_2 &= A_{0}C_{\frac23\pi}A_{0} ,\\
L_3 &= D_{-\frac12\pi} A_{0} A_{\frac12\pi} D_{0}  B_{\pm\frac12\pi} D_{0} A_{-\frac12\pi} A_{0}D_{\frac12\pi} ,\label{L3} \\
L_4 &=A_{\frac12\pi}A_{0}A_{-\frac12\pi}A_{0} A_{0}A_{-\frac12\pi}A_{0}A_{\frac12\pi} ,\label{L4}
\end{align}
\ese
where $D=\quarter \pi (1+\epsilon)$ is a nominal $\pi/4$ pulse.
One can easily notice that $L_1$ is equivalent to our type-1 sequence $\T_2^{(1)}$, Eq.~\eqref{T1-2}, while $L_2$ is equivalent to our type-2 sequence $\T_2^{(2)}$, Eq.~\eqref{T2-2}.
In Fig.~\ref{fig3}, the nine-pulse sequence $L_3$ is compared with our type-1 sequence $\T_3^{(1)}$, Eq.~\eqref{T1-3}, as they both have a total pulse area of $4\pi$.
Clearly, the profile of our sequence outperforms the profile of $L_3$.
Finally, the $L_4$ sequence delivers the same profile as our sequence  $\T_3^{(1)}$.

More generally, it can be shown that the profiles of Levitt's concatenated CPs, constructed by recursive expansion \cite{Levitt83}, coincide with our type-1 CPs.
However, the procedure described in \cite{Levitt83} is less flexible as it only allows for total pulse areas equal to $2^n\pi$, with $n=1,2,3,\ldots$. 

\sec{Conclusion}

In this paper we presented three classes of symmetric broadband composite pulse sequences, which compensate imperfections in the pulse area to an arbitrary high order.
The composite phases are given by simple rational multiples of $\pi$ for arbitrarily long sequences, and the transition probabilities for the three classes are given by very simple formulas.
These composite sequences perform equally well or outperform the existing composite sequences in the literature.
The current sets are more flexible, as they allow total pulse areas of arbitrary integer multiples of $\pi$.

\acknowledgments
This work has been supported by the Bulgarian Science Fund grant DN 18/14.


\appendix
\sec{Appendix}
In this appendix we will sketch the proof that the composite sequences of types 1, 2, and 3 lead to BB excitation profiles.

First, we take the product \eqref{Utot} for $T_N^{(1)}$ and after some simple trigonometry we obtain
\be\label{sumcos}
U_{11}^{(N)} = \sum_{j=0}^{N-1}\sin^{2j}\left( \frac{\pi\epsilon}{2} \right)Z_{j+1}(\phi) ,
\ee
where $\phi=\{\phi_k\}$ is the set of all composite phases and $Z_{j+1}$ do not depend on $\epsilon$.
We will only be interested in $Z_{N}$ and we will prove that it is equal to unity. Explicitly, we have
\be
Z_{N} = \half\e^{ -\i\left( \phi_1 + \phi_N + 2\sum_{j=2}^{N-1}\phi_j \right)}\prod_{j=1}^{N-1}\left( \e^{\i\phi_j}+\e^{\i\phi_{j+1}} \right)^2,
\ee
which, after substituting the explicit phases \eqref{phasesCP1} and applying some simple transformations can be written as
\be
Z_{N} = \half\e^{ -\i\left( N-1 \right)\pi/2}\prod_{j=1}^{N-1}\left( 1 + \e^{\i(2j-1)\frac{\pi}{2(N-1)}} \right)^2.
\ee
Finally, this can be rewritten as
\be\label{zn}
Z_N=\half 2^{2(N-1)}\prod_{j=1}^{N-1}\cos^2\frac{(2j-1)\pi}{4(N-1)}.
\ee
By using the property of the Chebishev polynomial of the first kind \cite{Zwillinger}
\be
T_n(x)=2^{n-1}\prod_{j=1}^{n}\left[ x-\cos \frac{(2j-1)\pi}{2n} \right],
\ee
one can prove that
\be
\prod_{j=1}^{n}\cos \frac{(2j-1)\pi}{4n} = \frac{\sqrt{2}}{2^n}.
\ee
This relation, when substituted in Eq.~\eqref{zn}, leads to
\be\label{Zn1}
Z_N=1 .
\ee
We now focus our attention back at Eq.~\eqref{sumcos}. Due to Eq.~\eqref{Zn1}, in order to fit into the required inequality $|U_{11}^N|<1$, we must have
\be
Z_{j+1}=0 \quad (\forall j \neq N-1).
\ee
Hence, we finally arrive at
\be
U_{11}^{(N)} = \sin^{2(N-1)}\left( \frac{\pi\epsilon}{2} \right)
\ee
and
\bse
\begin{align}
&P_{1\to 1}^{(N)} = \sin^{4(N-1)}\left( \frac{\pi\epsilon}{2} \right)\label{profile1-1} ,\\
&P_{1\to 2}^{(N)} = 1 - \sin^{4(N-1)}\left( \frac{\pi\epsilon}{2} \right) , \label{profile1-2}
\end{align}
\ese
which proves the BB condition for the set of phases \eqref{phasesCP1}.
In an identical way one can prove that for the CPs of types 2 and 3 we have
\be\label{profile2}
P_{1\to 2}^{(N)} = 1 - \sin^{4N-2}\left( \frac{\pi\epsilon}{2} \right).
\ee
If we take into account the total pulse area $A_{\text{tot}}$ of the CPs, we can unite Eqs.~\eqref{profile1-2} and \eqref{profile2} into
\be
P_{1\to 2} = 1-\sin^{2A_{\text{tot}}/\pi}\left( \frac{\pi\epsilon}{2} \right).
\ee


\end{document}